**Damage nucleation phenomena: Statistics of times to failure**

**(Short title: Annealed…)**

**S.G. Abaimov[1], A. Roy, and J.P. Cusumano[2]**

**Pennsylvania State University, Department of Engineering Science and Mechanics**

**[1]Electronic address: sgabaimov@gmail.com**

**[2]Electronic address: jpc3@psu.edu**

**Abstract.** In this paper we investigate the statistical behavior of an annealed continuous damage model. For different model variations we study distributions of times to failure and compare these results with the classical case of metastable nucleation in statistical mechanics. We show that our model has a tuning parameter which significantly determines the model behavior. Depending on the values of this tuning parameter, our model exhibits statistical behavior either similar to nucleation of systems in statistical mechanics or an absolutely different type of behavior intrinsic only for systems with damage. This lets us investigate the possible similarities and differences between damage phenomena and classical phenomena of nucleation in statistical mechanics.

## 1. Introduction

Damage as a complex phenomenon has been studied by many authors. A survey of recent developments in damage mechanics can be found in [1-3]. Studies recently appeared in the literature illustrate the similarity between damage phenomena and phenomena of phase transitions [4-15]. This similarity would give an opportunity to apply the well-developed formalism of statistical mechanics to the occurrence of damage. Therefore many attempts [5-7, 14-21] have been made to apply equilibrium statistical mechanics to damage phenomena. However, this question remains far from being completely resolved. The reason is that damage phenomena usually exhibit more complex behavior than gas-liquid or magnetic systems, and, in spite of what seems to be straightforward, the applicability of statistical mechanics to damage is subtle because the direct application can often cause the appearance of incorrect results [14, 15, 21].

Damage phenomena can generally be separated into two different categories. In the first category, damage behavior inherits thermal fluctuations from the medium in which it occurs (annealed behavior). The main representative of thermal damage fluctuations is the Griffith theory. The application of statistical mechanics here has many parallels with gas-liquid systems. In the second category, even in the case of non-thermal (quenched) systems the occurrence of damage has a complex behavior that can also be described by the formalism of statistical mechanics [14, 15]. However, all resulting equations in this case are valid not for energy characteristics of damage but for its topological properties. This type of behavior is often observed when the dynamical time scale of fracture is much faster than the time scale of thermal fluctuations and conductivity, so the dissipation processes have no time to attenuate the quenched disorder. In this case *a priori* input disorder in a model plays the crucial role.

In this paper we investigate the annealed behavior of a continuous damage model [22]. In Section 2 we introduce the model. In Section 3 we investigate the post-critical behavior of the model. In Section 4 we introduce stochastic noise into the system to simulate the irreversible pre-critical behavior for damage systems. In Section 5, to make a correspondence with the results of classical phenomena of nucleation, we investigate what happens if our model becomes reversible. In Section 6 we introduce a tuning parameter which switches the behavior of the system from the irreversible, 'damage' type to the reversible, 'classic' behavior in statistical mechanics. This let us illustrate the difference between damage phenomena and classical phenomena of nucleation.

## 2. Model

In this paper we utilize the continuous damage model developed by Cusumano *et al.* [22]. The model is used to simulate the mesoscale physics of elastic media and is based on the principle of action minimization of theoretical mechanics. In our simulations, a formulation with 128 finite elements is used. Numerical simulations are accomplished with a program based on the Open Source Library deal.II [23]. Further details on the model and action principle can be found in [22].

The evolution of the model follows the evolution of its displacement $u(t,x)$ and its damage $\varphi(t,x)$ in time-space domain until the model failure when at one of location the damage $\varphi$ reaches unity. Following [22], for the evolution of the model the non-dimensional equations are

$$\frac{\partial^2 u}{\partial t^2} - c\nabla^2 \frac{\partial u}{\partial t} - \nabla(1-\varphi)\nabla u = 0 \, , \tag{1a}$$

$$\frac{\partial \varphi}{\partial t} = \eta \left\langle \frac{\varphi}{2}(\nabla u)^2 - \alpha \varphi^{2/3} \right\rangle , \tag{1b}$$

$$u\big|_{x=0} = 0 \, , \ \left\{(1-\varphi)\nabla u\right\}\big|_{x=1} = F \, , \tag{1c}$$

where $<x> = x$ when $x \geq 0$ and zero otherwise, which makes our model irreversible when there is no healing and damage can only grow. The value of parameter $c$ dictates the amplitudes of damping processes in the system and the value of parameter $\alpha$ represent the damage threshold, above which the damage is allowed to grow. Parameter $\eta$ represents the time scale of damage growth.

One main difference of the model employed by Cusumano *et al.* [22] from that used for our simulations is that we utilize the constant load $F = 3 \cdot 10^{-4}$ as an external boundary constraint. We choose constant load for this publication for multiple reasons. First, the majority of results in statistical mechanics are for constant boundary constraints. Because one of the major goals of this study is to compare damage and classical phenomena of nucleation, to develop this comparison we use constant boundary constraints. Second, fatigue behavior of a system under an oscillating load is more complex than the constant load response. Therefore, we follow the principle of going from 'from simple to complex' and postpone the investigation of fatigue due to model oscillations for further studies.

## 3. Post-critical (supercritical) nucleation with the initial disorder

The growth rate of damage, given by Eq. (1b), is proportional to the difference between the tendency for the damage to grow $\frac{\varphi}{2}(\nabla u)^2$ and the threshold $\alpha \varphi^{2/3}$. This threshold works like a threshold in Griffith theory, or a potential barrier in nucleation. Only cracks above the critical value $\varphi_C = \left( \frac{2\alpha}{(\nabla u)^2} \right)^3$ are allowed to grow.

For the constant load $F = 3 \cdot 10^{-4}$, used in our simulation, the strain $\nabla u$ at earlier stages of damage growth is constant throughout the model and equals the load $F$. Following Cusumano *et al.* [22] for the choice of parameters we have $\alpha = 1.178 \cdot 10^{-11}$ and $\eta = 1.87 \cdot 10^3$. For this value of $\alpha$ the critical threshold of damage is $\varphi_C \propto 2 \cdot 10^{-11}$. For $c$ we use a larger value than [22] $c = 5$ to damp dynamical oscillations. For the same reason as the initial values of the displacement $u$ we use a static solution of Eq. (1a) for the given load.

For the initial values of damage, similar to [22], we utilize a uniform distribution in the range from 0 to 0.01, independently and identically distributed in the spatial variable $x$. For the critical threshold $\varphi_C \propto 2 \cdot 10^{-11}$ the initial damage at all locations is much higher than $\varphi_C$ with a probability close to unity. Therefore, simulations with these initial conditions correspond to a post-critical model of nucleation with the initial nuclei of another phase well above the critical nucleus size. Therefore, we expect a burst evolution of damage in simulations. Because the initial nuclei have sizes of the order of $10^{-4}$-$10^{-2}$ (as a uniform distribution from 0 to 0.01 among 128 elements) which are very much larger than critical value of $10^{-11}$, we expect this burst evolution to be so fast and so deterministic that neither variations of irreversibility criterion nor the introduction of possible thermal microfluctuations in the system would have any significant influence on the model behavior.

Since fluctuations are neglected, the growth of damage in the post-critical model starts from *a priori* defined initial values, and for each particular realization of initial damage follows a deterministic trajectory. To reach a rupture threshold, when at a particular location the value of damage becomes equal to unity, a finite time is required. The damage variable is present in the equation of 'interactions' (1a) only inside of the expression $(1-\varphi)$, which for small values of damage $\varphi \ll 1$ does not influence the evolution (1a) of $u$ almost until the point of the rupture. Therefore, during almost all of the time prior to rupture, all locations of the model are decoupled and growing their damage independently. Only at the latest stages of damage evolution does $\varphi$ become of the order of unity and non-linear effects of interactions among different locations start to influence the statistics. Therefore, because for small levels of damage the evolution of damage at a particular location is independent from the damage state of other locations, the site of rupture is determined *a priori* by a location with the maximal initial damage. Since for all simulations this location has the initial value of damage always close to 0.01 (as a maximum of elements of a sample distribution with the upper boundary 0.01), the times to failure are almost deterministic and have similar values for all simulations with a very small variance.

To apply trial distributions, we shift the statistics of times to failure by the minimum time to failure. The cumulative distribution function of shifted times to failure $t_f$ is given in Fig. (1a) as a cdf plot and in Fig. (1b) as an exponential plot (a specific plot for the exponential trial distribution where this distribution becomes a straight line). Also in these figures we plot the maximum likelihood fit of an exponential trial distribution. We see that the statistics of times to failure is Poissonian (exponential). To the extent of our knowledge we do not know experimental studies that would investigate the post-critical distribution of times to failure. However, if these studies were available, the best goodness-of-fit distribution would probably be different. The post-critical behavior is determined by non-linear, coupled effects of the behavior of a particular model. Therefore, it is not universal and is supposed to be specific for each particular material or structure.

## 4. Pre-critical (subcritical) irreversible nucleation

In the previous section we investigated nucleation of damage for the system with initial disorder. The sizes of nuclei in the model were much higher than critical. Therefore, as expected, we observed a burst evolution of damage in the model, when fluctuations play no role, and the behavior is deterministic and is determined by the initial, quenched disorder.

In this section we investigate another type of nucleation when initial nucleus sizes are much lower than the critical threshold. If in Section 3 we had the system 'rolling downwards' into a global minimum of a free energy potential, in this section we investigate the system 'climbing up' the potential barrier from the state of a local, metastable minimum of the free energy potential.

In the previous section the evolution of the system was deterministic and was determined by the initial distribution of nuclei. However, we should not employ the same method here. Indeed, any initial disorder well below the critical nucleus size will have no influence on the further system evolution, and will not result in the deterministic damage growth. Instead, dynamical thermal fluctuations must play the crucial role, and only when the size of these fluctuations has overwhelmed the potential barrier will the system burst. Therefore from this point on we do not use disorder in the initial conditions.

If we take a closer look at Eqs. (1), we see that Eq. (1a) is macroscopic and determines the evolution of macroscopic displacement $u$ in the system. On the other hand, Eq. (1b) is mesoscopic, determining damage evolution on the level of defects. As mesoscopic level here we refer not to particular cracks but to a crack density. The microscopic, atomistic level of singular defects is not present in our equations directly.

However, the implementation of microscopic dynamics is crucial for our simulations. Indeed, Eq. (1b) does not contain any possibility for the system to evolve below the critical threshold (growth rate is zero forever). Later, in following sections, we will relieve the criterion of irreversibility to compare our results with the theory of nucleation. In this case we see that the threshold returns the system to a state of zero damage. In other words, Eqs. (1) of our model are deterministic and do not support fluctuating behavior. However, this behavior is different from damage phenomena we observe in Nature: any solid has a constant process of birth and death of defects due to thermal fluctuations. That is, on the microscopic level thermal fluctuations can influence the system's behavior, exhibiting complex fluctuating interactions of damage and strain on this scale.

Eq. (1b) represents the mesoscopic mechanics of damage growth, and we need to introduce thermal fluctuations for its subcritical evolution. Each degree of freedom in statistical mechanics has averaged fluctuations $k_B T / 2$ because of the equipartition of energy. If we imagine a piston on a spring as a boundary constraint for a gas in a volume, the piston will have Gaussian microoscillations of its position, and its averaged energy will be $k_BT$ due to thermal fluctuations. In the same way, the neighborhood of any defect will have Gaussian microfluctuations of strain. For our model we introduce such fluctuations for the local strain as:

$$\nabla u(x.t) = \nabla u(x.t) + \Upsilon \xi(x,t).$$  (2)

where $\xi(x,t)$ is a Gaussian white noise with zero mean and unity standard deviation. A similar approach has been suggested in [24-27]. However, thermal fluctuations are microscopic and do not influence the mesoscopic level of Eq. (1b) directly. And of course they do not influence the macroscopic level of Eq. (1a). Therefore we do not include their influence into Eq. (1a) for the strain evolution and we should be careful when we are including them into Eq. (1b) for the damage evolution. If we would directly substitute $\nabla u$ in Eq. (1b) from Eq. (3), for the value of $\Upsilon$ we would have to use the microscopic constant of the order of $k_BT$. This noise would be negligible on the mesoscopic level and would have no influence on the damage evolution. This problem is well-known in damage mechanics, and experimental studies [28, 29] show that the variance of actual fluctuations is much higher than $k_BT$. Many authors [24-26, 29, 30] attributed this behavior to complex interactions of micro-disorder in a system (i.e., the presence of microdefects can cause the amplification of fluctuations). Another possible alternative is to associate this phenomenon with the influence of thermal fluctuations on the unstable, frustrated parts of defects, generally crack tips, on the microscopic level. Although these fluctuations are spatially and quantitatively microscopic, and influence only microscopic parts of cracks, their presence causes crack growth on the mesoscopic level. The 'sensitive' crack tip works in this case as an amplifier, causing the microscopic thermally-induced fluctuations to determine the mesoscopic crack growth. The third possible explanation is provided by considering a phenomenon observed in bubble chambers in particle physics. In that case, radiation of high-energy particles can facilitate nucleation [31] and cause the effective temperature to be higher than the 'actual' temperature of a specimen. This effect should be especially distinctive for the materials working in the conditions of high radiation. The suggestion for the radiation in normal environmental conditions to influence the growth of defects in solids requires experimental verifications. However, the counterpart of this effect for gas-liquid systems is well known and widely utilized in bubble chambers.

Therefore, in the mesoscopic Eq. (1b) we include the influence of the effective mesoscopic fluctuations of the strain with an amplitude that is much higher than the amplitude of thermal fluctuations. In other words, we substitute Eq. (3) into Eq. (1b) but with the fluctuations of strain that have the order of the strain by setting Y = 1:

$$\frac{\partial \varphi}{\partial t} = \eta \left\langle \frac{\varphi}{2} (\nabla u + Y \xi(x,t))^2 - \alpha \varphi^{2/3} \right\rangle. \tag{3}$$

However, we should modify Eq. (3) further. If we simulate the system well below the critical threshold, the probability of successful attempts to grow damage is expected to be small. In following section we will relieve the criterion of irreversibility. In this case there is a possibility that Eq. (3) attenuates damage to zero. And, if damage is zero, Eq. (3) does not contain any possibility for the system to evolve further, since the rate of damage growth has a power-law dependence on damage and is zero if damage is zero. Contrary to this, in Nature there is always non-zero level of micro-damage, as a result of fluctuations. Therefore we introduce a non-zero level of damage $\varphi_0$ below which the system cannot go: $\varphi = \max(\varphi, \varphi_0)$. In other words, at each time-step of our simulations we check to see if damage has fallen below the level $\varphi_0$ or not, and, if it does, we restore the damage back to the level $\varphi_0$.

Thus, finally, instead of Eq. (1b) we obtain

$$\frac{\partial \varphi}{\partial t} = \eta \left\langle \frac{\varphi}{2} (\nabla u + Y \xi(x,t))^2 - \alpha \varphi^{2/3} \right\rangle, \text{ where } \varphi \geq \varphi_0 \text{ always.} \tag{4}$$

A careful choice of parameters is required to provide a reasonable time of numerical simulations. We utilize $\varphi_0 = 1.5 \cdot 10^{-6}$, $\eta = 10^8$, and $\alpha = 2 \cdot 10^{-8}$. This high value of $\alpha$, which represents the energy cost of opening a crack's free surface, gives for the critical damage $\varphi_C = \left( \dfrac{2\alpha}{(\nabla u)^2} \right)^3 \propto 0.01$ for $\nabla u \propto F = 3 \cdot 10^{-4}$. This high, macroscopic value of the critical threshold provides that almost the total damage evolution from $1.5 \cdot 10^{-6}$ to $0.01$ is a pre-critical nucleation. Only after a long time, when the level of damage fluctuates above the critical threshold of $0.01$, does the evolution of damage switch from the pre-critical to post-critical regime, providing burst damage growth. However, the duration of the post-critical burst is very short in comparison with the long, 'random walking' time of many 'attempts' of the pre-critical fluctuation to exceed the threshold. Therefore our statistics of times to failure are almost pure statistics of the pre-critical fluctuating behavior.

The cumulative distribution function of (non-shifted) times to failure $t_f$ is given in Fig. (2a) as a cdf plot and in Fig. (2b) as a Weibull plot. Also in these figures we plot the maximum likelihood fits of Weibull and gamma trial distributions. We see that the statistics of times to failure is close to the Weibull distribution with exponent 1.83±0.02. However, Fig. (2b) shows that, although the statistics are close to Weibull, it is in fact a gamma distribution with exponent 2.85±0.05. This is an interesting result because it appears to contradict the fact that the Weibull distribution has been chosen to be a best-fit distribution in many previous studies of pre-critical damage nucleation [32-34]. To examine this subtle issue more thoroughly we have to implement additional verifying simulations.

As was discussed above, almost the total evolution time of the model is the fluctuating random walk of many successful or unsuccessful attempts to exceed the threshold, and only a negligible fraction of time is spent by the system in the final, burst, post-critical state before rupture. In Section 3, where we specifically studied the post-critical, burst stage of nucleation, during which the non-linear coupled interactions of different locations play a significant role, we were still able to qualitatively explain results by neglecting these interactions during almost all of the system evolution. In the current section, dealing with pre-critical nucleation, neglecting all possible interactions among different locations is an even better approximation, and we can thus integrate Eq. (4) as a separate ordinary differential equation for each independent location.

However, we should remember the general principle of damage mechanics that rupture is a 'horse race' among different locations, and the time to failure is the time of the first 'winner'. Therefore, initially we construct a statistics of all resulting times to failure, as if all elements were loaded independently (later we will refer to these statistics as 'min of 1' statistics, that is, as the statistics of one particular model element). As a second step, we group consecutive independent results of Eq. (4) in groups of 100 and choose minimum time to failure in all groups. This corresponds to the rupture of a model or a solid built of 100 independent elements. Therefore, we henceforth refer to these statistics as 'min of 100' statistics. In a similar way we investigate the rupture of solids consisting of 500, 1000, and 5000 elements to construct 'min of 500', 'min of 1000', and 'min of 5000' statistics respectively.

The cumulative distribution functions of times to failure $t_f$ are given in Fig. (3a) as cdf plots and in Fig. (3b) as Weibull plots. Also in these figures we plot the maximum likelihood fits of Weibull and gamma trial distributions for the 'min of 1' and 'min of 5000' statistics. We see that for the 'min of 1' statistics, which are the statistics of failure times of a single element, the distribution of failure times is gamma with exponent 8.78±0.12. The Weibull distribution for this statistics is clearly not applicable. However, when we increase the number of elements in the model (when we move from the 'min of 1' statistics to 'min of 5000' statistics), the sample distribution step-by-step transforms from the gamma distribution to the Weibull distribution, and for the 'min of 5000' statistics we obtain already a good fit of the Weibull distribution with exponent 20.4±0.2. For the continuous model in Fig. (2) we had 128 elements in the model. Therefore we can conclude that our results for the continuous model in Fig. (2) represent an intermediate stage of the transfer process when the gamma distribution is still valid but the Weibull distribution becomes valid. For the general case we can conclude that the statistics of times to failure for one particular, undivided element in our model is the gamma distribution while in the thermodynamic limit of an infinite number of elements the statistics approaches the Weibull distribution. Therefore we see, as expected, that in applications, all finite element models deviate from the thermodynamic limit of the Weibull distribution due to a finite-size effect. The number of elements they utilize should be determined by the required accuracy of engineering simulations.

The important fact here is that this distribution is gamma or Weibull but *not* exponential, and therefore our results do not correspond to classical phenomena of nucleation theory. The appearance of non-exponential distributions in nucleation has been previously found to take place in systems with amorphous disorder, when a free energy potential has multiple minima, and has been suggested for the cases of polymer crystallization [35] and glass-forming materials [36]. The primary difference of our model from classical gas-liquid nucleating systems is the irreversibility of damage. Due to fluctuations the system can exceed the threshold of $\alpha$ to grow damage further but cannot decrease the density of cracks already present in the system. Therefore natural to expect that particularly the irreversibility causes our results to diverge from the nucleation theory. In the next section we will turn our attention to the case of a reversible system.

## 5. Reversible nucleation

In Section 4 we followed [22] and imposed the condition of damage irreversibility, as indicatedby the operator $<x>$ in Eq. (4). However, the majority of studies in the theory of nucleation investigate reversible systems. Indeed, in the theory of gas-liquid systems, if a small bubble of another phase appears in a metastable state, there is no constraint for these systems that would prohibit to this bubble from disappearing. The same is true for magnetic systems where nothing prohibits a small domain of another phase from disappearing. To compare our results with previous studies in nucleation theory, in this section we will relieve the condition of irreversibility and will allow defects to disappear. In other words, instead of

Eq. (1b), in this section above the damage level $\varphi_0$ we allow negative damage growth rates ('healing') via

$$\frac{\partial \varphi}{\partial t} = \eta \left( \frac{\varphi}{2} (\nabla u + \Upsilon \xi(x,t))^2 - \alpha \varphi^{2/3} \right), \text{ where } \varphi \geq \varphi_0 \text{ always,} \qquad (5)$$

*without* the angle brackets <...>. Also in this section, to provide a reasonable time for numerical simulations, we utilize a higher value of the microdamage base level, $\varphi_0 = 0.0003$, which is still well-below the critical value $\varphi_C = \left( \frac{2\alpha}{(\nabla u)^2} \right)^3 \propto 0.01$. Other parameters we keep unchanged. Again, damage evolution from 0.0003 to 0.01 dominates the duration of the simulations up to failure, and our statistics of times to failure are almost pure statistics of pre-critical fluctuating behavior.

The cumulative distribution function of (non-shifted) times to failure $t_f$ is given in Fig. (4a) as a cdf plot and in Fig. (4b) as an exponential plot. Also in these figures we plot the maximum likelihood fit of an exponential trial distribution. We see that the statistics of times to failure is Poissonian (exponential). This result is similar to the result obtained by Bonn et al. [37] and also to results of nucleation theory [31, 38, 39].

Similar to the previous section we verify our results with decoupled numerical simulations. The cumulative distribution functions of times to failure $t_f$ are given in Fig. (5a) as cdf plots and in Fig. (5b) as Weibull plots for 'min of 1', 'min of 100', 'min of 500', 'min of 1000', and 'min of 5000' statistics. Also in these figures we plot the maximum likelihood fits of exponential trial distributions. In Fig. (5a), to exhibit all data on a single plot, we rescaled times of failure 80, 320, 510, 1600 times for 'min of 100', 'min of 500', 'min of 1000', and 'min of 5000' statistics respectively. For the same reason we utilized in Fig. (5b) the Weibull plot instead of more appropriate exponential plot. We see that the statistics of times to failure are Poissonian (exponential).

The exponential statistics of times to failure in this case is expected. A brittle solid ruptures as well as a liquid nucleates when the size of fluctuations overwhelms the critical, activation energy. For our model the specific activation energy is

$$E_C = \alpha \varphi_C^{2/3} = \varphi_C (\nabla u)^2 / 2 = \frac{4\alpha^3}{(\nabla u)^4} \qquad (6)$$

The probability for a fluctuation to reach energy level $E$ is

$$p(E) \propto \exp(-const \cdot E / Y^2); \qquad (7)$$

therefore times to failure are distributed exponentially and the averaged time to failure [40] is proportional to

$$t_f \propto \exp(const \cdot E_C / Y^2) \propto \exp\left(const / (\nabla u)^4\right). \qquad (8)$$

Here $\nabla u$ is the strain in the model. However, almost the whole duration of the pre-critical damage nucleation takes place when the damage is low in the model (fluctuations in the vicinity of $\varphi_0 = 0.0003$) and does not influence Eq. (1a) of the stress redistribution. Therefore, all pre-critical nucleation does not distinguish between constant stress and constant strain as possible boundary constraints, and we can use the external force $F$ instead of the strain $\nabla u$ in Eq. (8):

$$t_f \propto \exp\left(const / F^4\right). \tag{8}$$

We see that the logarithm of the averaged time to failure is inversely proportional to the fourth power of the constant external strain or constant external stress as a boundary constraint. This is a direct consequence of the Griffith theory [41-43]. Similar results of load dependence were obtained experimentally by Guarino et al. [29, 44] for irreversible wood and fiberglass. Pauchard and Meunier [28] obtained similar dependence for two-dimensional solids with the inverse proportionality to the second power of strain/stress

$$t_f \propto \exp\left(const / F^2\right). \tag{9}$$

Dependence (9) was also found in numerical investigations of a fiber-bundle model with noise [24]. As it was discussed by Bonn et al. [37], the general dependence for the averaged time to failure is

$$t_f \propto \exp\left(const / (\nabla u)^\tau\right) \propto \exp\left(const / F^\tau\right) \tag{9}$$

where the exponent $\tau$ is determined by the dimensionality of a system and by the fractality of the structure of microdisorder. Our model provides $\tau = 4$.

## 6. Partial reversibility

In previous sections we investigated two extreme case of the complete irreversibility, intrinsic for brittle materials, and the complete reversibility, intrinsic to liquids and gels. In this section we consider an intermediate case of partial reversibility.

As it was suggested by Golubovic and Feng [42], Golubovic and Peredera [45], processes of surface and body diffusion can relieve the stress in the crack's neighborhood which was cause by the crack formation. This leads to the conclusion that the longer a particular part of a crack exists, the less it becomes reversible. As a simplest dependence for the reversibility we assume that a given fraction $D$ of damage is reversible. In other words, if $\varphi_{\max}$ is the maximum value of damage that occurred so far at a given location, for this location we assume that damage is reversible in the range $(1 - D)\,\varphi_{\max}$ to $\varphi_{\max}$ and is irreversible in the range 0 to $(1 - D)\,\varphi_{\max}$. This choice seems to be reasonable. If we consider an ellipsoidal crack, then the condition that faction $D$ of damage is reversible is equivalent to the condition that the fraction $R = 1 - \sqrt{1 - D}$ of crack radius is reversible while the fraction $1 - R = \sqrt{1 - D}$ of crack radius is irreversible. The same condition that reversible is a fraction of crack radius was originally used by Golubovic and Feng [42], Golubovic and Peredera [45].

As examples we consider the cases of partial reversibility $D = 25\%$, $D = 50\%$, $D = 75\%$. The cumulative distribution functions of times to failure $t_f$ are given in Figs. (6a,c,e) as cdf plots and in Fig. (6b,d,f) as Weibull plots. Also in these figures we plot the maximum likelihood fits of Weibull and gamma trial distributions. We see that the statistics of times to failure is again in a transfer state from the gamma distributions to the Weibull distributions.

Again, to illustrate system's behavior, we compare results with the decoupled model. As an example we consider the case $D = 50\%$, in other words half-reversibility of damage. The cumulative distribution functions of times to failure $t_f$ for the decoupled model are given in Fig. (7a) as cdf plots and in Fig. (7b) as Weibull plots. Also in these figures we plot the maximum likelihood fits of Weibull and gamma trial distributions for the 'min of 1' and 'min of 5000' statistics. We see that for the 'min of 1' statistics, which are the statistics of failure times of a single element, the distribution of failure times is gamma with exponent 7.11±0.12. The Weibull distribution for this statistics is clearly not applicable. However, when we increase the number of elements in the model (when we move from the 'min of 1' statistics to 'min of 5000' statistics), the sample distribution step-by-step transforms from the gamma distribution to the Weibull distribution, and for the 'min of 5000' statistics we obtain already a good fit of the Weibull distribution with exponent 2.85±0.03. First, we see that the behavior of partial reversibility is more similar to the completely irreversible case than to the completely reversible. However, the Weibull exponent is much lower for this case. Therefore for systems with restricted reversibility we expect the behavior to be different from nucleation theory of completely reversible systems. For the general of partial reversibility case we can conclude that the statistics of times to failure for one particular, undivided element in our model is the gamma distribution while in the thermodynamic limit of an infinite number of elements the statistics approaches the Weibull distribution.

## 7. Conclusions

In our study we investigate the behavior of damage nucleation. Particularly, we concentrate on the statistics of times to failure. We consider two distinctive cases of the post-critical, burst nucleation, when the system has overwhelmed already the potential barrier and of the pre-critical, 'random walk' nucleation, when the system 'climbs up' the potential barrier by means of large fluctuations. For the last case of subcritical nucleation we discover the reversibility of damage significantly determines damage behavior. So, for the reversible case we repeat results of nucleation theory in gas-liquid systems while for the irreversible or partially reversible case we obtain the Weibull distribution for failure times. This study indicates that damage phenomena represent a specific type of nucleation phenomena with many own intrinsic features, and caution should be executed while nucleation theory is applied to the case of damage.

**Acknowledgements**

This work was partially supported by a grant from Siemens Corporation.

**Figures:**

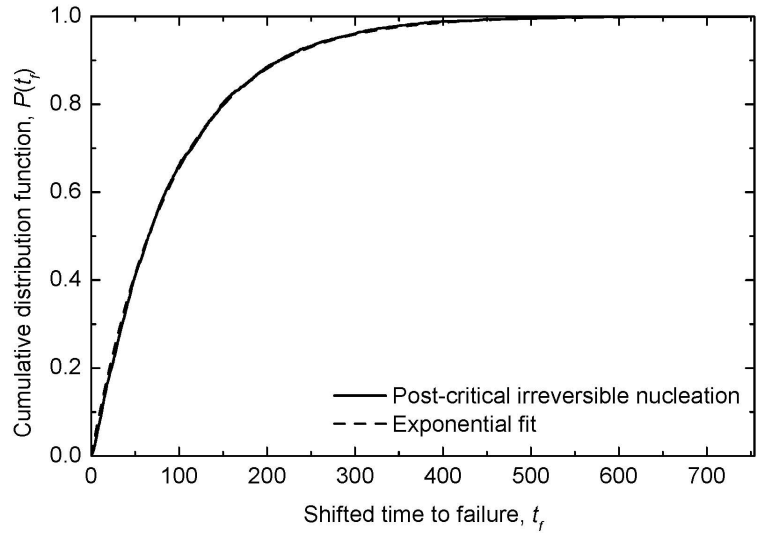

(a)

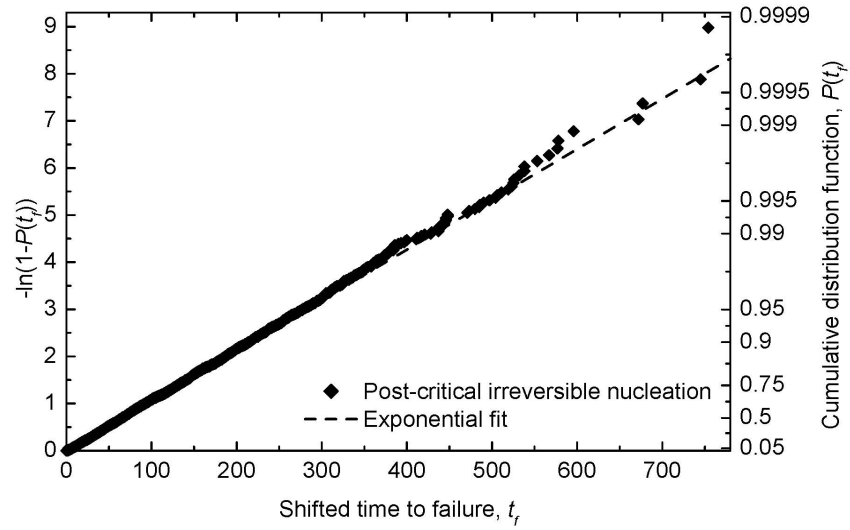

(b)

Figure 1. Cumulative distribution function of shifted times to failure $t_f$ for the post-critical irreversible nucleation, (a) cdf plot and (b) exponential plot.

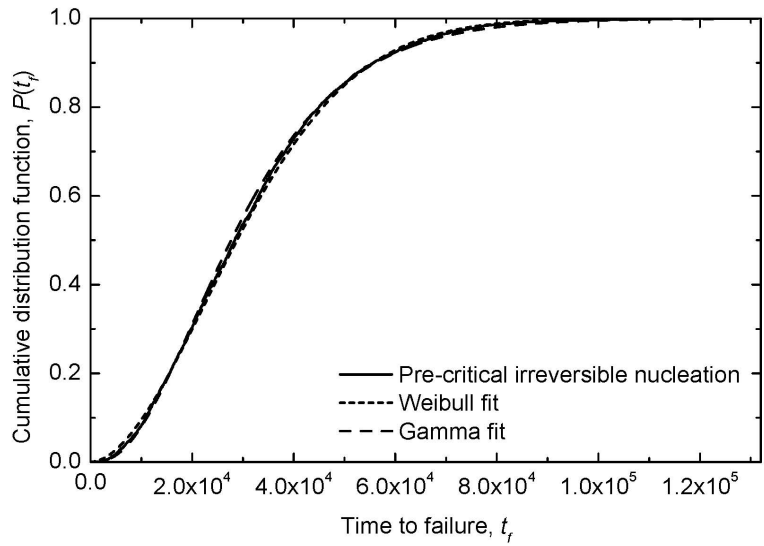

(a)

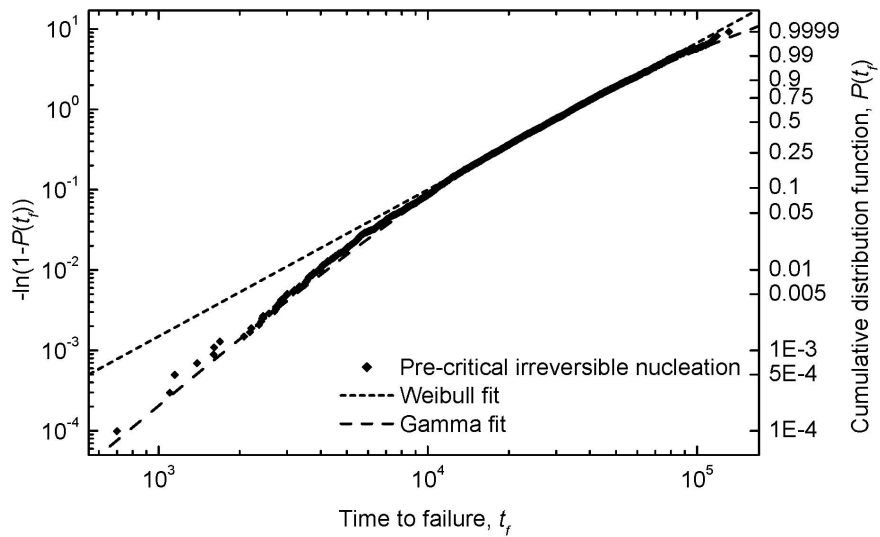

(b)

Figure 2. Cumulative distribution function of times to failure $t_f$ for the pre-critical irreversible nucleation, (a) cdf plot and (b) Weibull plot.

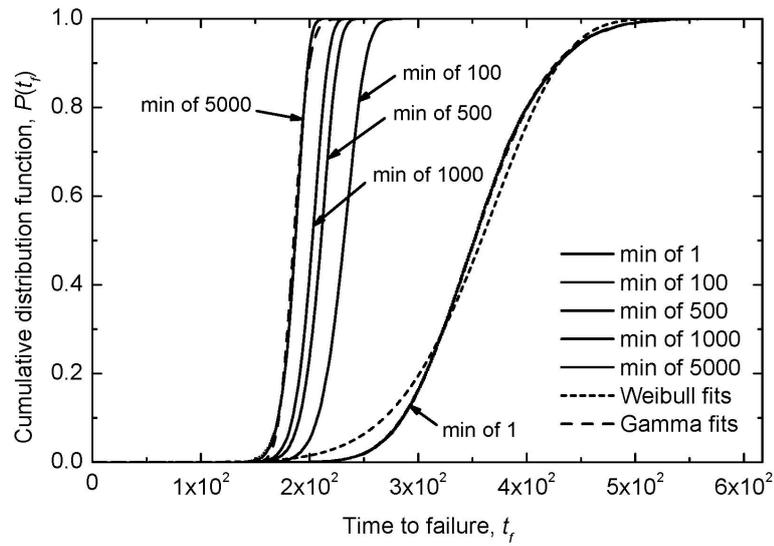

(a)

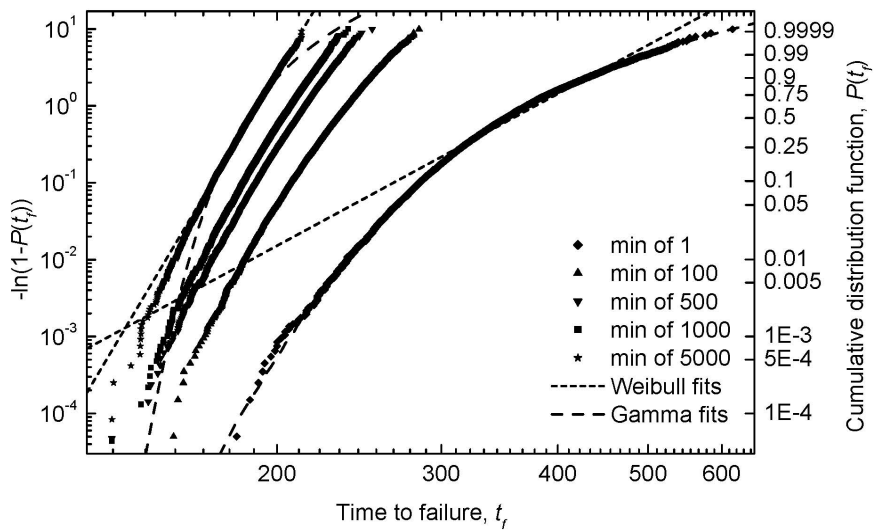

(b)

Figure 3. Cumulative distribution function of times to failure $t_f$ for the pre-critical irreversible nucleation of independent locations, (a) cdf plot and (b) Weibull plot.

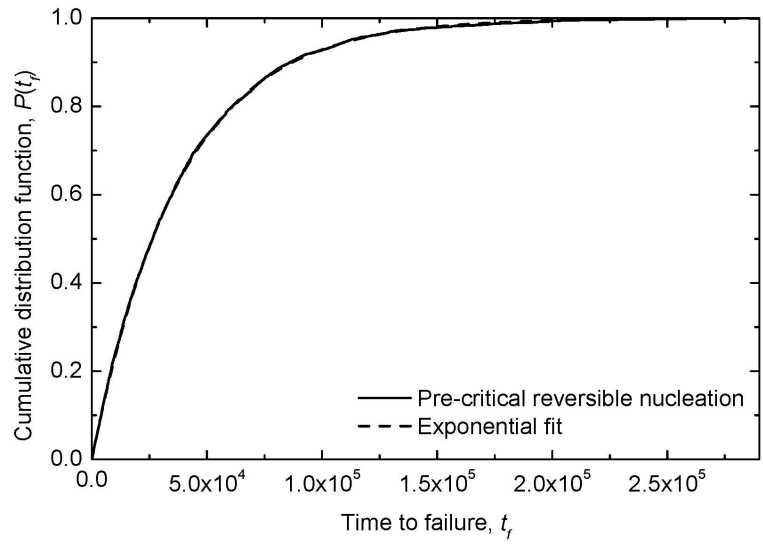

(a)

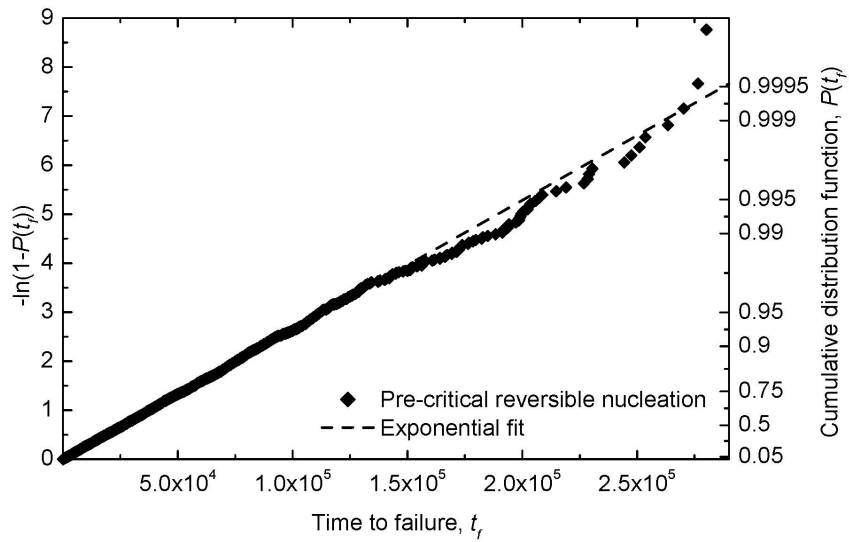

(b)

Figure 4. Cumulative distribution function of times to failure $t_f$ for the pre-critical reversible nucleation, (a) cdf plot and (b) exponential plot.

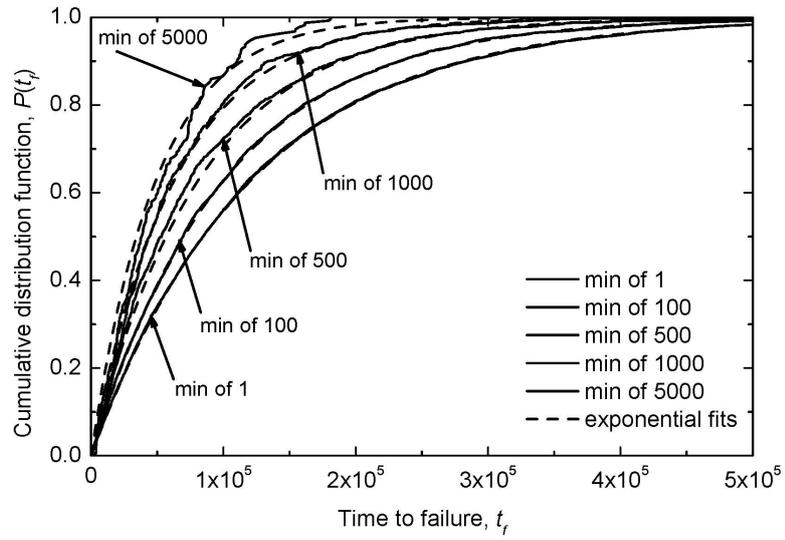

(a)

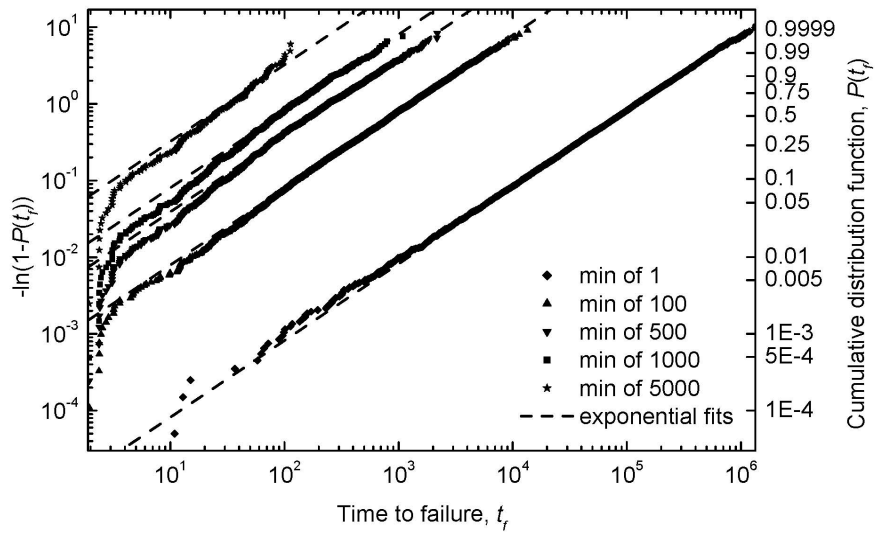

(b)

Figure 5. Cumulative distribution function of times to failure $t_f$ for the pre-critical reversible nucleation of independent locations, (a) cdf plot and (b) Weibull plot. In (a) we rescaled times to failure 80, 320, 510, and 1600 times

for the 'min of 100', 'min of 500', 'min of 1000', and 'min of 5000' statistics respectively.

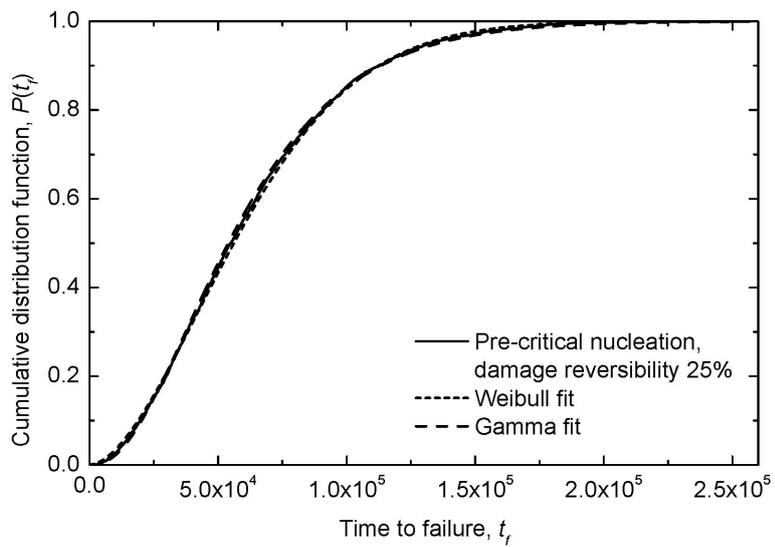

(a)

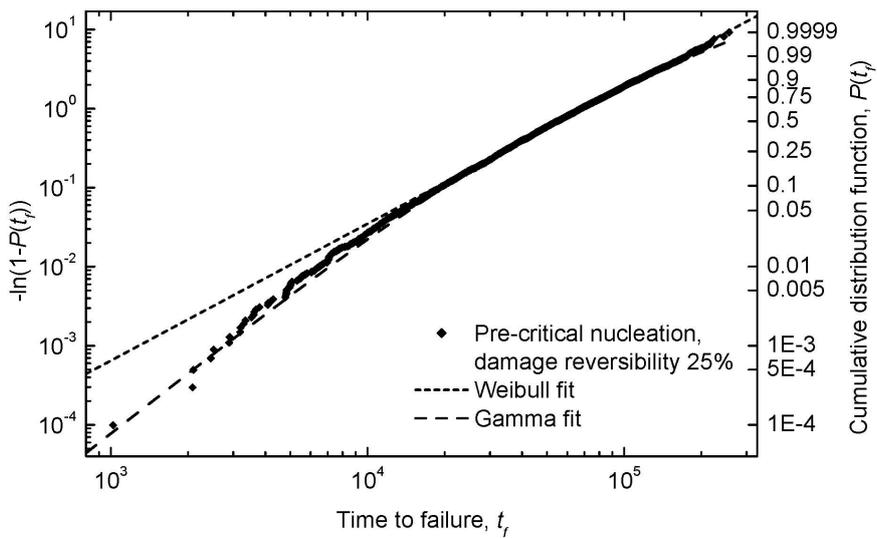

(b)

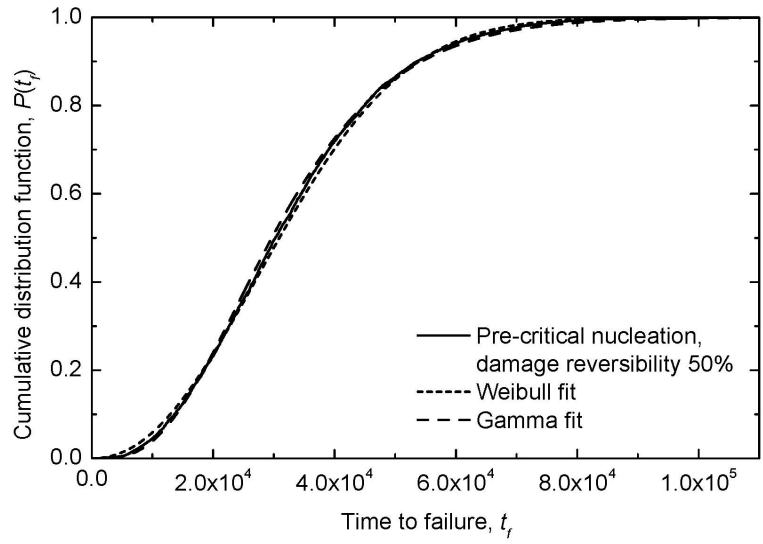

(c)

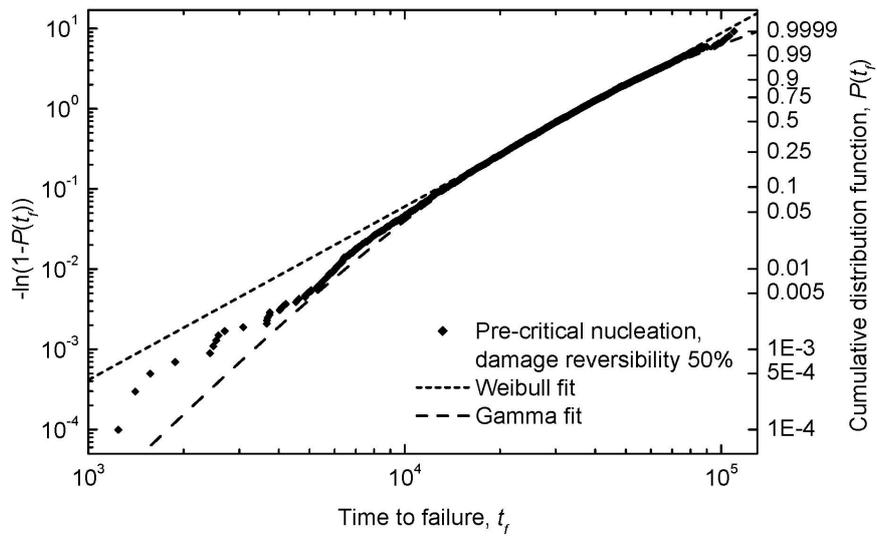

(d)

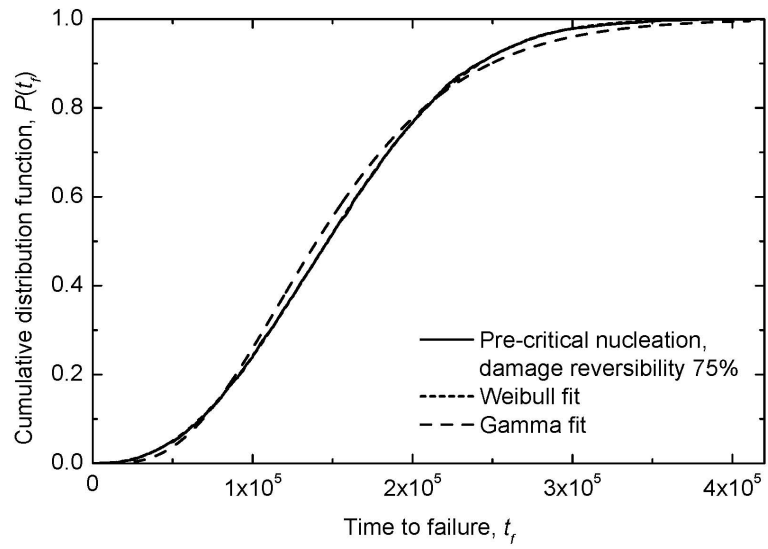

(e)

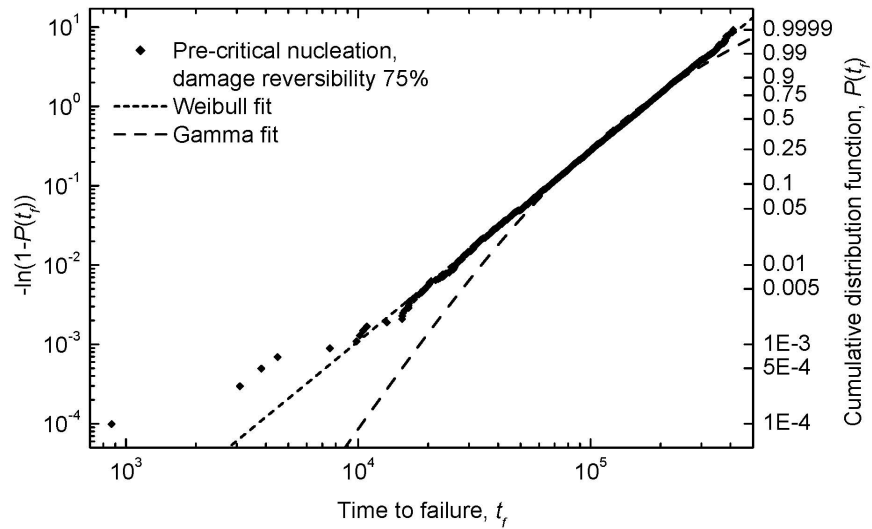

(f)

Figure 6. Cumulative distribution function of times to failure $t_f$ for the pre-critical nucleation with damage reversibility $D = 25\%$, (a) cdf plot and (b) Weibull plot, $D = 50\%$, (c) cdf plot and (d) Weibull plot, $D = 75\%$, (e) cdf plot and (f) Weibull plot,.